\documentclass[conference]{IEEEtran}
\IEEEoverridecommandlockouts
\usepackage{booktabs} 
\usepackage{url}
\usepackage{color}
\usepackage{soul}
\usepackage{booktabs}
\usepackage{graphicx}
\usepackage{multirow}
\usepackage{mathrsfs}
\usepackage{amssymb}
\usepackage{amsbsy}
\usepackage{amsmath}
\usepackage{tcolorbox}
\usepackage{listings}
\usepackage{stackengine}
\usepackage{algorithm}
\usepackage{algpseudocode}
\usepackage{esvect}

\begin{document}

\title{Towards effective AI-powered agile project management}


\author{ 
\IEEEauthorblockN{Hoa Khanh Dam}
\IEEEauthorblockA{University of Wollongong, Australia \\ {\tt \small hoa@uow.edu.au}}
\and
\IEEEauthorblockN{Truyen Tran} \IEEEauthorblockA{Deakin University, Australia \\ {\tt \small truyen.tran@deakin.edu.au}}
\and
\IEEEauthorblockN{John Grundy}
\IEEEauthorblockA{Monash University, Australia \\ {\tt \small john.grundy@monash.edu}}
\and
\hspace{3cm}
\IEEEauthorblockN{Aditya Ghose}

\IEEEauthorblockA{\hspace{3cm} University of Wollongong, Australia \\ 
\hspace{3cm} {\tt \small aditya@uow.edu.au}}

\and
\IEEEauthorblockN{Yasutaka Kamei}
\IEEEauthorblockA{Kyushu University, Japan \\ {\tt \small kamei@ait.kyushu-u.ac.jp}}

}

\maketitle


\begin{abstract}
The rise of Artificial intelligence (AI) has the potential to significantly transform the practice of project management. Project management has a large socio-technical element with many uncertainties arising from variability in human aspects e.g., customers' needs, developers' performance and team dynamics. AI can assist project managers and team members by automating repetitive, high-volume tasks to enable project analytics for estimation and risk prediction, providing actionable recommendations, and even making decisions. AI is potentially a game changer for project management in helping to accelerate productivity and increase project success rates. In this paper, we propose a framework where AI technologies can be leveraged to offer support for managing agile projects, which have become increasingly popular in the industry.
\end{abstract}

\begin{IEEEkeywords}
Software engineering, artificial intelligence, agile project management
\end{IEEEkeywords}

\section{Introduction}

Artificial Intelligence (AI) has started making a substantial impact to many parts of our society, and is predicted to disrupt  how we produce, manufacture, and deliver. The rise of AI is empowered by the growth and availability of big data, breakthroughs in AI algorithms (e.g. deep learning), and significantly increased computational power. The pervasiveness of software products has resulted in a massive amount of data about software projects which AI techniques can leverage. We envision that AI will transform (software) project management practice in many aspects, from automating basic administration tasks to delivering analytics-driven risk predictions and estimation, facilitating project planning and making actionable recommendations. In this paper, we present a framework of how various AI technologies are adapted and integrated to support various areas of agile project management (\emph{agile PM}).

Agile methods (e.g. Scrum) have been widely used in industry to manage software projects \cite{hoda2018}. This relatively new approach to  project management empowers software teams to focus on rapid delivery of business value to customers, thus significantly reducing the overall risk of project failures. Project management has thus witnessed a shift away from the traditional ``waterfall'' process and towards a more adaptive, agile model. The number of projects following agile has increased significantly in the recent years, not only in the software industry but also in other non-IT domains \cite{StateofAgile}.


An agile project are centered around a \emph{product backlog}, which is typically a collection of items to be completed in the project \cite{Cohn2005}. Items  in a product backlog can be, for example, customer requirements for the product (user stories), requests for bug fixes, changes to existing features, and technical improvements. Product backlog is evolved through regular updates and refinement to ensure that it contains items that are relevant the project's scope and objectives, sufficiently detailed, and appropriately estimated. Important updates to the product backlog include adding or removing backlog items based on current needs, estimating the size of items, and refine large items into small fine-grained items.


\begin{figure}[ht]
\centering \includegraphics[width=\linewidth]{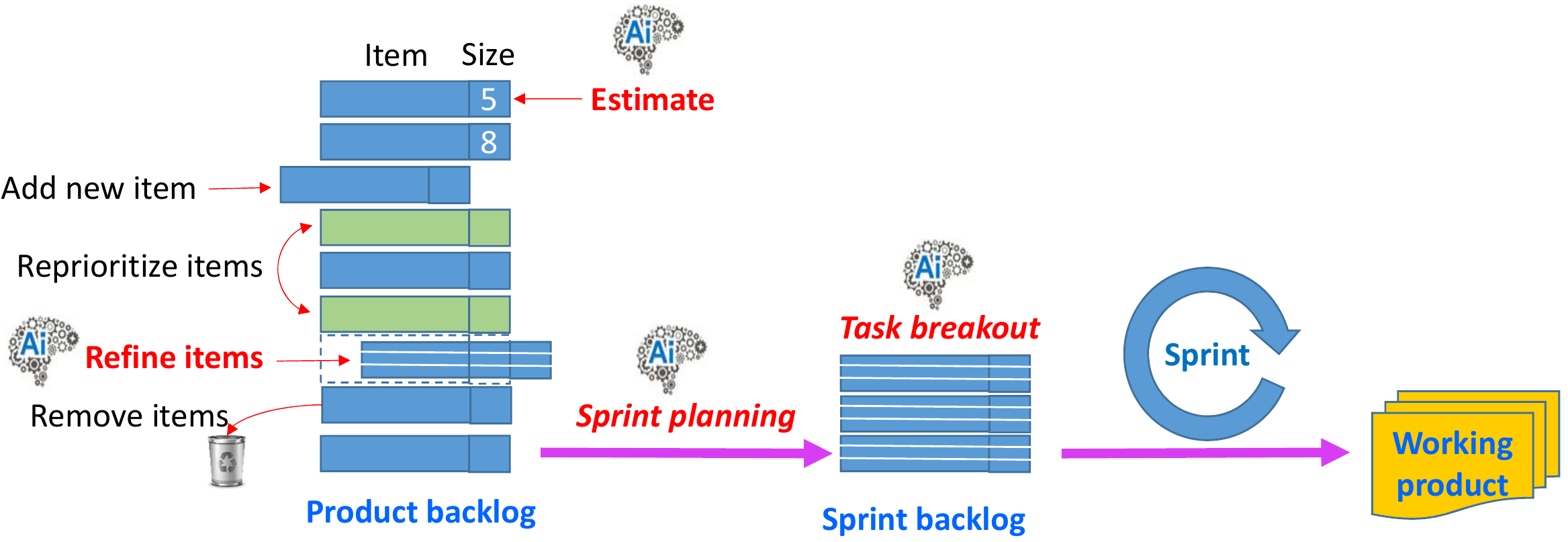}
\caption{A typical agile process}
\label{fig:agile-process}
\end{figure}

An agile project consists of multiple iterations (or alternatively referred as \emph{sprints}). Each sprint is often a short period in which the team aims to complete a subset of items in the product backlog. Prior to a sprint, the team performs \emph{sprint planning} to identify the goal of that upcoming sprint, and select items from the product backlog which they will complete to meet the sprint's goal. During sprint planning, many agile teams decompose each product backlog items into a set of \emph{tasks}. These tasks and their corresponding product backlog items form the \emph{sprint backlog}. The team then executes the sprint to complete items in the sprint backlog to deliver a potentially shippable product increment.


\section{Challenges in Agile Project Management}


Many tools have been developed to support agile project management such as Atlassian's JIRA Software\footnote{\url{https://www.atlassian.com/software/jira}}, Axosoft\footnote{\url{https://www.axosoft.com}}, and Assembla\footnote{\url{https://www.assembla.com}}. Those tools allow agile teams to create and manage various agile artifacts such as user stories, product backlogs, sprints, and sprint backlogs. For example, they support teams in creating user stories and tasks, linking related tasks and user stories, assigning team members to tasks and issues, creating deadlines, setting priorities and estimates (e.g. story points). They also enable team members to see the amount of work required for individuals and teams during each sprint, track progress of the sprint and its associated user stories and tasks. They facilitate real-time information exchange and collaboration via centralised project information.

Although existing agile tools are useful, their support is limited to creating, managing, and tracking project artifacts, and visualising historical project data such as burndown charts and other agile reports. Current agile project management tools lack advanced analytical methods that are capable of harvesting valuable insights from project data for prediction, estimation, planning and action recommendation. Many decision-making tasks in agile projects are still performed by agile teams without machinery support. We identify a number of important areas in agile project management that remain challenging due to this lack of effective support.

\subsubsection{\textbf{Identifying backlog items}} Items in the product backlog can be derived from different sources such as a requirement specification, new feature requests from customers, bugs reported by end users, previous bug fixes, discussions among agile teams (e.g. technical debts, design changes or action items from retrospective meetings), end users' reviews of the product, and even experiences from previous projects.  It is difficult and time consuming for agile teams, especially product owners, to process this large amount of heterogenous data in order to identify and create new items for the product backlog. In addition, for each newly created backlog item, it is necessary to consider \emph{inter-dependencies} between the new item and existing ones. This is challenging as a typical project has a large product backlog with more than 100 items.

\subsubsection{\textbf{Refining backlog items}}  Some items (e.g. user stories) in the product backlog are initially large, thus do not fit within a single sprint. Agile teams are often required to refine these large items into small ones such that they not only facilitate implementation but are sufficiently large, allowing stakeholders to understand business value \cite{Cohn2005}. There are typically three levels of refinement: (1) decomposing an epic into a number of user stories; (2)  splitting user stories into small stories; and (3) breaking a user story into a number of specific project tasks. Different rules and guidelines have been proposed to help teams refine backlog items, but rules often overlap with or even conflict with one another. Teams struggle to refine backlog items and rely on their own intuition and experience.

\subsubsection{\textbf{Sprint planning}} The key part of sprint planning is selecting a subset of items in the backlog which can realistically be accomplished by the team in the upcoming sprint to deliver a product increment. The customer expects the team to deliver what have been planned for a sprint, thus meeting this expectation is important in maintaining the customer's faith in the team's ability to deliver. Sprint planning is however highly challenging since many important factors must be considered, including items contributing toward the sprint goal, their priority and business value to customers, the dependencies among items, appropriate allocations to bug fixing and other technical work (e.g. resolving technical debts) and the availability of team members and the team's capacity. Risks impeding a sprint execution should also be forecasted and factored into a sprint plan. Sprint planning thus requires not only in-depth understanding of the current project and team but also experience learned from previous projects. Tool support is needed to manage complexities for large projects.

\subsubsection{\textbf{Pro-actively monitoring sprint progress and managing risks}} As the sprint unfolds, the team needs to track sprint progress and manage risks. Current practices in risk management mostly rely on high-level guidance and subjective judgements.  Predicting future risks is highly challenging due to the inherent uncertainty, temporal dependencies, and especially the dynamic nature of software. There is currently a gap in providing agile teams with insightful and actionable information about the current existence of risks in a sprint, and recommending concrete measures to deal with those risks.




\section{An AI--Powered Agile Project Assistant}

\begin{figure}[ht]
\centering \includegraphics[width=\linewidth]{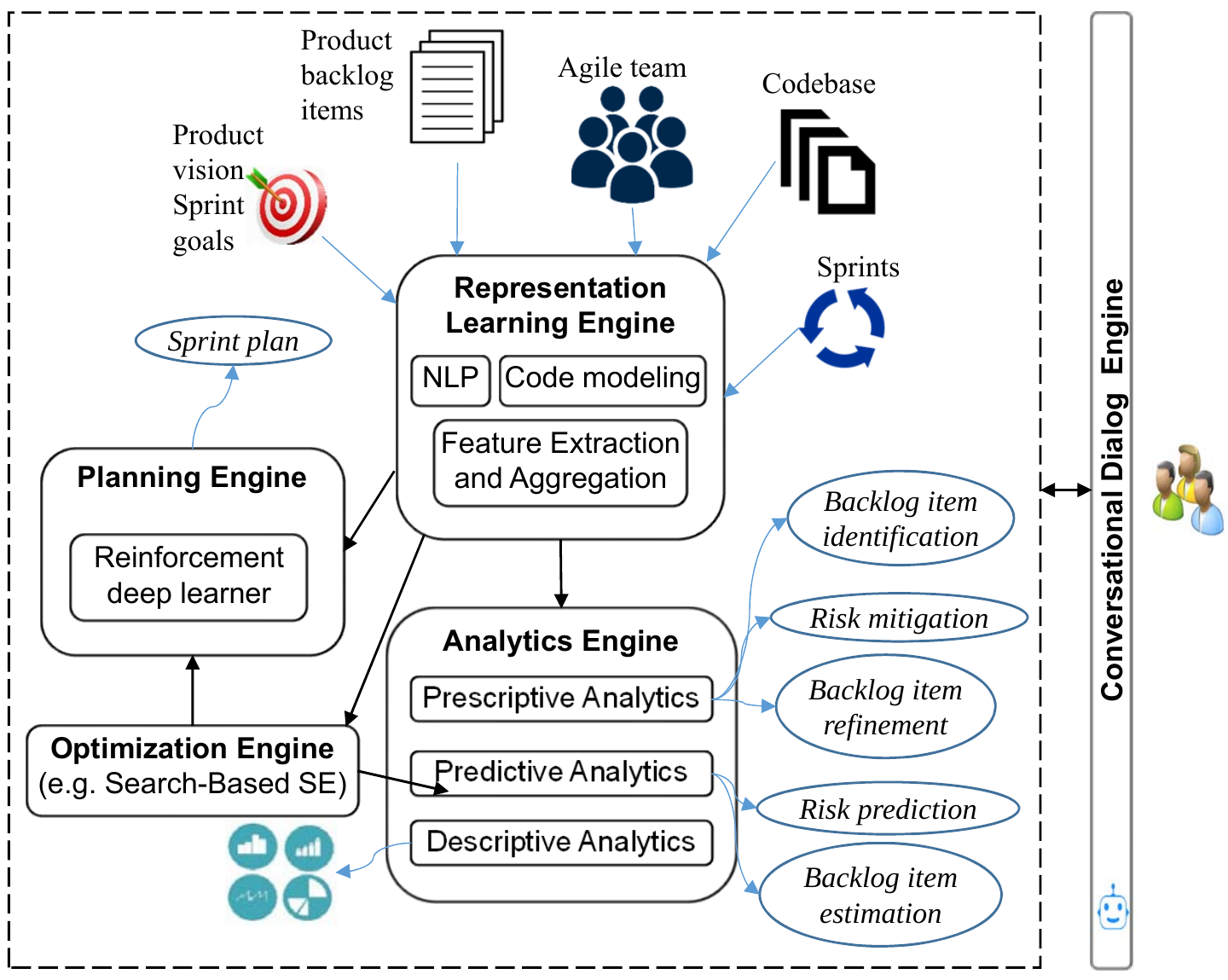}
\caption{The architecture of an AI-powered agile project management assistant}
\label{fig:framework}
\end{figure}

The above challenges and the serious lack of effective tools presents an opportunity for AI to significantly improve the practice of agile project management. AI-based tools are able to process massive amounts of data generated from software projects, harvest useful insights, and train to perform complex tasks such as estimating effort, task refinement, resource management, and sprint planning. Figure \ref{fig:framework} shows our proposal for the architecture of an AI-powered agile project management assistant. The core of this AI system are an \emph{\textbf{analytics engine}}, a \emph{\textbf{planning engine}} and an \emph{\textbf{optimization engine}}. These machineries depends on the \emph{\textbf{learning representation engine}} to learn and generate representations of project data that are mathematically and computationally convenient to process. The \emph{\textbf{conversational dialog engine}} converses with users and brings the support provided by the other engines to the users.

\subsection{Representation learning  engine}
Agile project artifacts contain both structured and unstructured data. For example, backlog items may have structured attributes such as type and priority (which are easily extracted to form a vector representation), whereas product visions, sprint goals, description of backlog items, and communication among team members (e.g. comments on backlog items) are written in natural text. Codebases contain documentations such as release notes and comments written in natural text, and source code written in programming languages. Hence, the representation learning engine is an important component of this AI system, responsible for learning meaningful vector representations for each project artifact. These representations can automatically be learned from unlabelled data, and are then used by the other machineries in the AI system.

The representation learning engine has a \texttt{NLP} component which performs automatic analysis on  project textual artifacts and then generates good representations of those artifacts. Traditional NLP techniques (e.g. Bag of Words) produce very high dimensional and sparse vector representations. By contrast, latest advances in deep learning-based NLP techniques \cite{manning2016computational} such as word2vec, paragraph2vec, Long Short-Term Memory (used in Google Translate), or Convolutional Neural Networks (used in Facebook's DeepText engine) are able to generate dense vector representations that produce superior results on various NLP tasks. Source code is another important source of project data. The \texttt{Code Modeling} component is responsible for learning meaningful vector representations which reflect the semantic and syntactic structure of source code. State-of-the-art statistical language modeling techniques, including deep learning models, have demonstrated  their effectiveness for source code and thus can be leveraged here \cite{allamanis2018survey,Tufano:2018:DLS}.

We have leveraged the powerful deep learning architecture, Long Short-Term Memory (LSTM) to automatically learn vector representations for both backlog items and source code\footnote{References omitted due to double-blind review requirement.}. LSTM enables us to learn the semantics and syntactic structures, particularly the long-term dependencies, existing in both natural text and source code. We will extend these models to learn representations for other textual artifacts such as product visions, sprint goals, and developer communications.

Any useful AI machinery must take into account the capability and dynamics of agile teams. Obtaining a representation for a team requires modeling of its members (e.g. developers). A developer can be represented through the project artifacts they have involved with, such as the backlog items they have completed or the code they have written. The \texttt{Feature Extraction and Aggregation} extracts all the vector representations of the artifacts related to a developer, and learn to aggregate them to form a vector representation of the developer. A number of feature aggregation techniques proposed in recent work  \cite{Choetkiertikul-TSE-2018} which derive vector features of a sprints based on the features of the backlog items assigned to it. We will extend those aggregation methods to learn features for representing team members. This representation will be enriched with features representing work and social dependencies between team members, extracted from communication logs (e.g. comments or discussions on work items).

\subsection{Analytics engine}

The analytics engine aims to provide decision support in the following aspects:

\subsubsection{\textbf{Descriptive analytics}} Most existing agile project management tools support this basic level of analytics: data visualization via reports, dashboards, and scorecards. Common agile reports such as burndown charts, velocity charts, and sprint reports are created by summarizing what happened using historical project data and presented to the users in an intuitive and easily interpretable manner. Knowing what happened (e.g. team velocity from sprint to sprint) is useful, but \emph{diagnosing} why something happened (e.g. why the team's velocity dropped significantly in some sprints) is even more useful. AI equipped with machine learning can augment descriptive analytics by discovering patterns, identifying anomalies and detecting ``unusual'' events.

\subsubsection{\textbf{Predictive analytics}} Most existing agile tools are not yet capable of providing this advanced level of analytics. Two challenging areas are effort estimation and risk prediction that are specifically for agile contexts. Machine learning techniques are suited to build prediction models. For example, recent work \cite{Choetkiertikul-TSE-2019} used deep learning to estimate the size of user stories through learning a team's previous estimates. Estimation tools could be used as a decision support system and takes part in the existing estimation process (e.g. planning
poker) or in a completely automated manner. Forecasting future risks requires the capability of processing large amounts of historical project data, memorizing a long history of past experience, and inferring the current ``health'' state of the project. Recent work has moved forwards in this direction to predict delay risks \cite{Choetkiertikul:2017:PDI} or sprint delivery risks \cite{Choetkiertikul-TSE-2018}.


\subsubsection{\textbf{Prescriptive analytics}} This is the most advanced level in the project analytics stack. Using the results from descriptive analytics and predictive analytics, prescriptive analytics recommends the best course of actions for agile teams in a specific situation. We identify here three important areas in agile PM that prescriptive analytics would be useful:
\begin{itemize}
  \item \emph{Backlog item identification:} Using the NLP component in the representation learning engine, prescriptive analytics will automatically process and extract new backlog items from  different data sources such as a requirement specification, new feature requests from customers, bugs reported by end users, previous bug fixes, discussions among agile teams (e.g. technical debts, design changes or action items from retrospective meetings), end users' reviews of the product, and even experiences from previous projects. It will also able to recommend inter-dependencies between new item and existing ones using machine learning and representation learning.

  \item \emph{Backlog item refinement:} prescriptive analytics will suggest how a user story is split into a smaller user stories or how a user story is decomposed into tasks. Learning decompositions is highly challenging since it requires a background knowledge. It is
      still a new topic in AI and machine learning.

  \item \emph{Risk mitigation:} Using results from predictive analytics, prescriptive analytics recommends a course of actions to take advantage of a future opportunity or mitigate a future risk and shows the implication of each decision option.
\end{itemize}

\subsection{Reasoning capability}

Reasoning is the capacity to infer new knowledge by algebraically manipulating existing knowledge base to respond to a query \cite{bottou2014machine}. Traditionally, it works on symbolic knowledge representation through the means of induction or deduction. This permits domain knowledge provided by agile teams (e.g. project rules). Recently, deep neural reasoning offers an alternative for producing answers from sub-symbolic (vector) representation \cite{jaeger2016artificial}, which are output of the representation engine. Inferred knowledge can be put back to enrich the knowledge base. The reasoning capability of our AI system is provided by two engines: planning and optimization.

\subsubsection{Planning engine}
Planning for a sprint can be formulated as an AI planning problem in which the initial state is the state of the project and the product prior a sprint, a goal state is specified in the sprint's goal, and selecting items from the product backlog can be viewed as the act of choosing plan operators to be executed from the initial state to a goal state.  The planning engine needs to consider a range of input such as the existing product backlog items, the sprint's goal, the existing codebase, the team's capacity and previous performance in previous sprints, and the duration of the sprint. These data are often not formally expressed. The representation learning engine has convert them into vector representations but further formal encoding would be needed.


In addition, the plan needs to be executed in a manner that is \emph{robust} and \emph{resilient} to changes. The challenge is not only to be flexible enough to deal with immediate impediments to the sprint execution, but to also anticipate future states of affairs that might impede sprint execution or the achievement of the sprint goal. Impediments to the successful sprint execution can appear in many forms. For instance, a task might not be completed by the due date, preventing other dependent tasks from being started. Hence, the relationship between a sprint plan and its
operating environment can been seen as adversarial. Recent successful work in deep reinforcement learning (e.g. \cite{mnih2015human}) can be thus leveraged to build this part of the planning engine.


\subsubsection{Optimization engine}
The optimization engine helps the planning engine to compute the optimal set of actions given a certain situation. For example, it can be used to compute the optimal selection of backlog items for the upcoming sprint given multiple constraints and objectives. It can also be used for hyper-parameters tuning of machine learning models used in the analytics engine. Search-based software engineering techniques can be leveraged here to build the optimization engine.

\subsection{Conversational dialog engine}
The conversation dialog engine is envisioned to converse meaningfully with agile teams. It is a form of a software chatbot \cite{Lebeuf2018}, acting as an interface between the users and the remaining part of the AI system. The chatbot can be asked different types of questions, such as ``Show me your estimate of this user story'' or ``Can you help split this user story?''. Through conversations with the users, it receives input and requests, and passes them to relevant engines in the system. Future chatbots can be trained end-to-end \cite{serban2016building} and person-specific instead of task-specific \cite{kottur2017exploring}.






\section{Next Steps}


We are developing prototype tools to realize each component of the proposed AI-powered agile project management assistant. We plan to first evaluate it using our existing dataset of 150 open source projects.  We will also collaborate with our existing industry partners to perform an evaluation on commercial software agile projects. We however believe that AI will assist, not substitute, human teams. Individuals, interactions, and collaboration are still the key elements of project success. AI can serve as a distinctive accelerator for agile teams and thus help increase project success rates.

\bibliographystyle{IEEEtran}

\end{document}